\documentclass{aa}
\usepackage{graphics}

\begin{document}
\thesaurus{03               
		(03.13.2;   
                 03.13.4;   
		 08.01.1;   
		 08.01.3;   
		 08.06.3)}  

\title{On-line determination of stellar atmospheric parameters $T_\mathrm{eff}$, 
$\log g$, [Fe/H] from ELODIE echelle spectra \\
I - The Method}

\thanks{based on observations made on the 193cm telescope at Observatoire
de Haute-Provence, France}

\author{D. Katz\inst{1}, C. Soubiran\inst{2}, R. Cayrel\inst{1}, M. Adda\inst{3},
R. Cautain\inst{4}}

\offprints{D. Katz, david.katz@obspm.fr}

\institute{Observatoire de Paris, DASGAL, 61 av. de l'Observatoire, F-75014 Paris, France \and
Observatoire de Bordeaux, BP89, F-33270 Floirac, France \and
Observatoire de Paris-Meudon, DESPA, F-92195 Meudon Cedex, France \and
Observatoire de Haute-Provence, F-04870 Saint-Michel l'Observatoire, France} 

\date{Received ; accepted }

\titlerunning{On-line determination of atmospheric parmeters}
\maketitle

\begin{abstract}

We present a method estimating the atmospheric parameters $T_\mathrm{eff}$, $\log g$, [Fe/H]
for stars observed, even at low signal to noise ratio, with the echelle spectrograph
ELODIE on the 193cm telescope at Observatoire de Haute-Provence.
The method relies on the least-square comparison of the spectrum of a target star 
to a library of 211 spectra
of reference stars for which the atmospheric parameters are well known and
which were observed with the same instrument.
In order to obtain a meaningful comparison between target and reference spectra,
all features which are not intrinsic to the objects must be removed.
This treatment involves the correction of the blaze efficiency for each order,
cosmic rays hits and telluric line removal, 
convolution of the spectra to a common spectral resolution,
wavelength scale and flux level adjustment.
The library available at the present time
covers the effective temperature range [4000K, 6300K], the metallicity range [$-$2.9, +0.35]
and the gravities of both unevolved and evolved stars existing at these temperatures
and metallicities.
Tests performed with the actual library allow us to estimate the internal accuracy to be
86 K, 0.28 dex and 0.16 dex for $T_\mathrm{eff}$, $\log g$, [Fe/H]
for a target star with S/N = 100
and 102 K, 0.29 dex and 0.17 dex at S/N = 10.
This accuracy will improve in the future
as the number of reference stars in the library will increase. 
The software (named TGMET) has been installed
at Observatoire de Haute-Provence for the on-line analysis 
of the high-resolution spectra of ELODIE, which was originaly conceived
for accurate radial velocity measurements.
                                      

\end{abstract}


\section{Introduction}
   The availability of immediate, on-line reduction for radial velocities at the
ELODIE echelle-spectrograph of the Observatoire de Haute-Provence, led to the  
obvious idea that it would be very useful to extend this type of fast analysis
to other stellar parameters.
Being ourselves involved in the determination of the metallicity of stars of known
proper motions in two galactic directions (Soubiran 1992, Ohja et al. 1994,
Perrin et al. 1995)
we were tempted to try to determine on-line the metallicity, 
effective temperature and gravity of a star observed with ELODIE.
Our former experience with this type of determination from spectra at a lower resolution
from a grid of synthetic spectra ( Cayrel et al. 1991a \& b, Perrin et al. 1995)
allowed us to
estimate that for old stars (with slow rotation) of solar type or cooler, a mean accuracy
of 200 K in $T_\mathrm{eff}$, 0.4 dex in gravity and 0.3 dex in metallicity was
possible from a spectrum obtained with a signal to noise ratio (S/N)
of 50 on a limited spectral interval. From theoretical considerations (Cayrel 1991)
we thought that with the higher resolution and larger spectral interval
of ELODIE's spectra, it would be possible to obtain comparable or better precision
on low S/N ($\sim$ 10) spectra.\\

We could have followed the same approach as before, and tried to compare the observed spectrum
with a grid of synthetic spectra, generated for example with the ATLAS9, SYNTHE
(Kurucz 1993) codes made generously 
available to us by R.L.  Kurucz. However we thought that there was a simpler method, i.e. comparing
the spectrum of the target star with a library of spectra of reference stars, with accurately
determined parameters, taken with the same spectrograph. The disadvantage is that an error
on the parameters of  reference stars  affects the result. Also the optimal extraction
of the parameters possible with a grid of synthetic spectra, where  the sensitivity
of any spectral feature to the parameters is known explicitely, is lost. The advantage is that
a very large spectral range can be used, without the very tedious work of fine-tuning
the oscillator strengths and damping constants of an extremely large number of atomic or
molecular lines. This approach would also avoid the empirical corrections
of synthetic to reference star spectra that have proven to be necessary (Cuisinier et al. 1994).\\
   
The principle sounds very simple. But, before the target spectrum can be meaningfully 
compared to the spectra of the library, many steps are required.
First of all it is necessary to remove all features which are not 
specific to the object, but instrumental in nature, or associated to particular conditions
of observation (Sect. 3). The main instrumental feature
is the modulation of the spectra by the blaze profile of each order. If the two objects
to be compared had the same radial velocity with respect to the instrument, at the time of
the exposures, this modulation would cancel out. But most of the time there is a significant
difference, and the blaze efficiency is shifted in wavelength between the two exposures.
This modulation must be corrected (Sect. 3.2).
Cosmic rays are a big nuisance in all instruments using CCDs as detectors. A first treatment
is made in the radial-velocity software (Baranne et al. 1996), following  Horne's algorithm (1986). Unfortunately
many cosmics escape the trap, because too few pixels are illuminated along the direction perpendicular 
to the dispersion. Therefore the remaining cosmics must be chased (Sect. 3.3).    
Equally disturbing are the telluric lines which change in intensity and position with time
in the rest frame of the object.
The pixels affected by these wandering disturbers must be eliminated (Sect. 3.6).
If the night is spoiled by the Moon, it may be
necessary to subtract a sky exposure, for which all the steps already described 
(except telluric line removal) must be
carried out too (Sect. 3.4).\\

After these different steps, three other actions remain to be performed. 
Two spectra to be compared must be brought :\\

\noindent
(i) $\ \ $ to the same spectral resolution\\
(ii) $\ $ to a common wavelength scale\\ 
(iii)$\ $ to a common level of flux\\

If the target star and a reference star of the library have a different line-broadening, because
they have a different projected rotational velocity $v\sin i$, we must not be fooled
into considering 
them as objects of different effective temperature, gravity and metallicity. Also it is not
guaranteed that the instrumental resolution is exactly the same for all observing runs
(e.g. because of focus variations). Therefore all the spectra of the library and the
target's star are forced to a common resolution (action (i) listed above),
in order to eliminate spectral differences
originating from projected rotation, macroturbulence or instrumental resolution
(Sect. 3.5).\\
 
Action (ii) is easy to perform because 
the radial velocity of the target star and of each comparison star is accuratly known from the
radial velocity software (Baranne et al. 1996), by cross-correlation with a mask
(Sect.  4.1). Action (iii) is done by least squares (Sect. 4.2) .\\

A large fraction of our observing time was devoted to the aquisition of the library of
reference stars, which includes 211 spectra at the present time. Sect. 2 describes the
observational material.
A detailed description of the
library is available in a companion paper (Soubiran et al. 1998, hereafter paper II).
The spectra of this library are available at CDS of Strasbourg, together with
their revised atmospheric parameters (Sect. 5).\\

Different tests have been performed to evaluate the consistency and the accuracy of
the method (Sect. 6).\\

The software is named TGMET, for Temperature, Gravity, METallicity.

\section{ELODIE's spectra}

ELODIE is a fibre-fed spectrograph devoted to the measurements of accurate radial 
velocities (Baranne et al. 1996). It has been in operation since 1993 on the 193cm 
telescope of Observatoire de Haute-Provence. The most striking result of this
instrument is the discovery of the first jovian planet orbiting the solar-type star
51 Peg (Mayor \& Queloz 1995). Its resolution power is 42000, the spectra range from
390 nm to 680 nm and are recorded  as 67 orders (numbered from 91 to 157)
on a $1024 \times 1024$ Techtronic's CCD. The
reduction of the spectra is automatically performed on-line, as well as the computation
of radial velocities by cross-correlation
thanks to the ELODIE reduction software developed by D. Queloz (1996). The precision of radial
velocities is about 100 m.s$^{-1}$ for F, G and K stars. A typical S/N of 100 is achieved with
one hour exposure on a $8.5^{th}$ magnitude star or a S/N of 10 on a $12.75^{th}$
magnitude star, taking into account a read noise of 8 e$^{-}$ per CCD pixel
or 15 e$^{-}$ per spectral resolution element. The different configurations of ELODIE
are very stable, allowing to compare easily spectra observed at different epochs.\\

To develop this method, we have observed, between 1994 and 1997, 250 reference 
stars at a mean S/N of 100, 211 of which
remained in the final library. We have also observed at low S/N (4 to 30)
30 well-known stars with 
various atmospheric
parameters to test the software. In parallel, we have achieved our
primary objective which was to
probe the Galactic stellar populations in two directions,
where we observed 132 stars of our astrometric program (Soubiran 1993, Ojha et al. 1994). 
Stars exhibiting a double correlation peak (identifying themselves as spectroscopic
binaries) during the radial velocity determination, were removed from the set of observations.
We have worked on already extracted spectra, 
i.e. on one-dimensional spectra, of 1024 spectral elements per order. We will call pixel
any of these spectral elements along a given order, although it is not a CCD pixel anymore.\\

\section{Preparation of the spectra}
\subsection{Wavelength calibration}
The extraction of the orders and the calculation of the coefficients of
interpolation for the wavelength calibration are  performed, just after the exposure, by
the ELODIE reduction software. These coefficients are used in formula $(1)$ to establish the
relation between pixels and wavelengths.

\begin{equation}
\lambda(x,k) = {1 \over k}\sum_{i=0}^3\sum_{j=0}^5
A(i,j)T_{i}(\tilde{x})T_{j}(\tilde{k}^{-1})
\end{equation}

where $k \in [91, 157]$ is the order of the spectrum, $x \in [1, 1024]$ 
the number of the pixel in a
given order, $T_{i}$ and $T_{j}$ the
values of the Chebyshev polynomial of order $i$ and $j$ at the points $\tilde{x}$
and $\tilde{k}^{-1}$ and $A(i,j)$ the
coefficients of the interpolation.
In equation $(1)$, $\tilde{x}$ and $\tilde{k}^{-1}$ are x and 1/k normalized
on the $[-1,1]$ interval by a linear transformation.

\subsection{Flat field correction and straightening of each order}
The ELODIE reduction software provides a flat field correction,
separated in two components. The first one is the response
correction, that takes into account pixel to pixel sensitivity differencies.
The extracted spectra we are using underwent this correction. The second one
is a division of each order by the response of the tungsten lamp spectrum of the same order,
to correct the spectra from the modulation of the blaze profile.\\

We decided not to use the same
procedure to straighten the orders, because of the presence of a coloured
filter in the beam. This
filter is necessary to balance the exposures at the red end and the violet
end of the tungsten lamp.
But it implies that the spectrum is not flat along the wavelength coordinate.
Therefore, polynomials of order 7 to 19 were fitted to the continuum of
four very metal-poor (almost line-free)
stars : \object{HD 221170}, [Fe/H] = $-$2.10,
\object{HD 216143}, [Fe/H] = $-$2.15, \object{HD 108317},
[Fe/H] = $-$2.36, \object{HD 140283}, [Fe/H] = $-$2.53 (Soubiran et al. 1998, paper II).
For each order $k$, the response
curve $B_k(x)$ is given by the polynomial which is less affected by the spectral lines,
with respect to the three others.
Instead of using the tungsten response, the observed spectra are straightened by
dividing their flux,
order by order, pixel by pixel, by $B_k(x)$, normalized at 1 at its maximum.
For further uses all the pixels that correspond to a response value lower
than $0.5$ at the edges of the orders are rejected by giving them the flag value $-100.0$.\\

Orders $138$ to $157$ (the bluest spectral orders) were rejected, because
they were underilluminated with respect to the rest of the orders, 
and  too much degraded by noise.\\

\begin{figure}[h]
\resizebox{8cm}{!}{\includegraphics{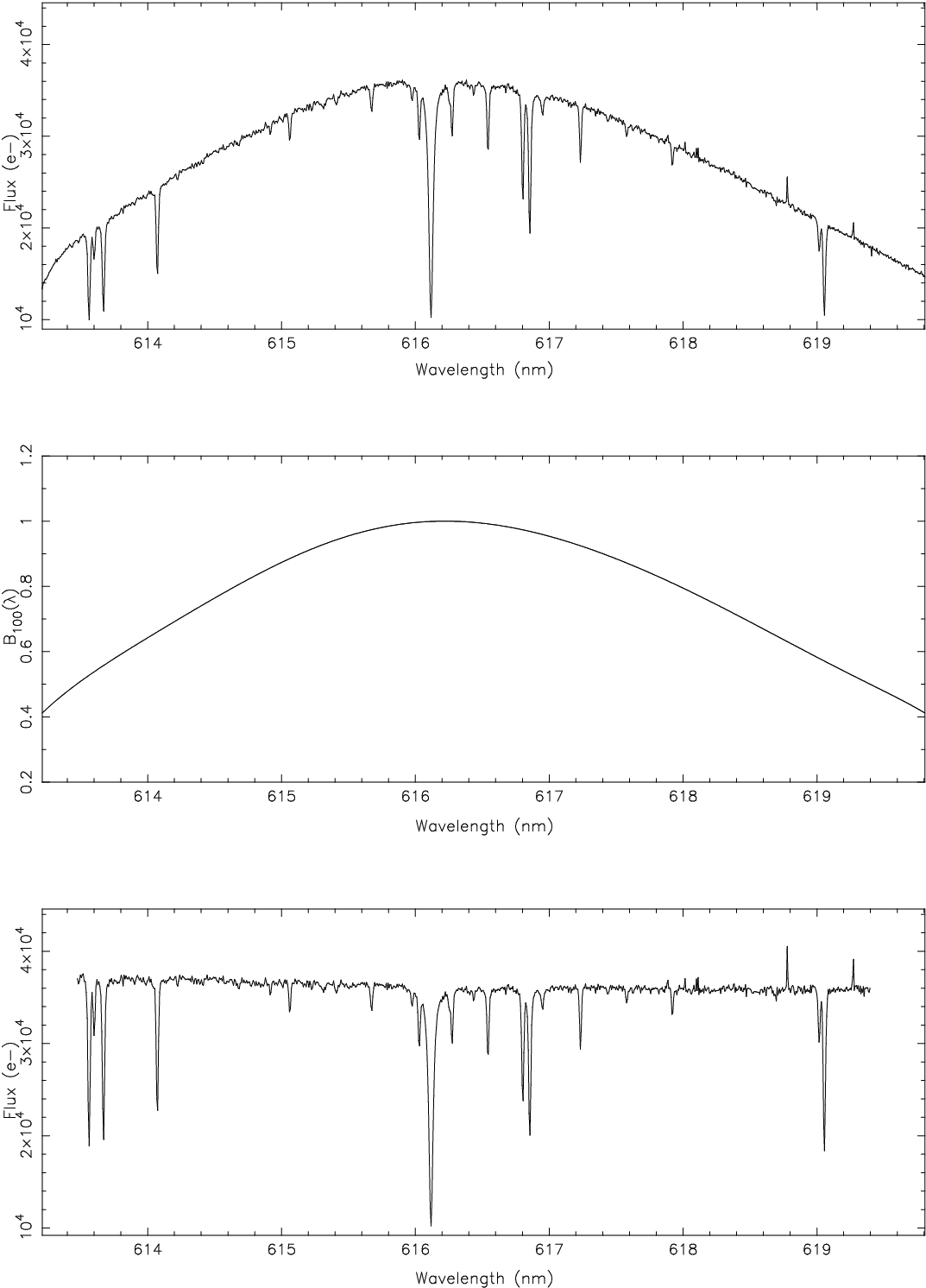}}
\caption{Order 100 of \object{HD 25329} ($T_\mathrm{eff}$ = 4787 K, $\log g$ = 4.58,
[Fe/H] = $-$1.72, S/N = 141) before (top) and after straightening (bottom) and
the response curve $B_{100} (\lambda)$ which has been used to straighten the order (middle).}
\label{}
\end{figure}

If the spectral energy distribution  (SED) of
the target star is very similar to that of the star which was used for straightening
the order, no significant slope in the SED of the straightened order is expected.
If the two stars have different colours, a residual slope is expected. 
Initially, the adjustement of the slope between the target and a reference star was
performed in the least square fit of the flux levels.
However a better consistency was attained in removing this adjustment. In practice the
slope remains very small
over an order: about 1\% from edge to edge for a 1200 K difference in effective
temperature. The ``successful'' comparison star has of course a temperature difference 
considerably smaller than this, with respect to the target star,
so the slope adjustment is not necessary.
Indeed it is even unwise, because if slope adjustment is provided, the software try to
compensate the presence of a strong line at the edge of an order in a comparison star, 
by inventing an unphysical strong slope in the target star, when the line is non-existent
or much weaker in the target star (or vice versa). Figure 1 shows order 100
of \object{HD 25329} before and after straightening.

\subsection{Cosmic rays and defective pixels removal}
The ELODIE reduction software corrects the 2D frames from
the cosmic rays (high energy particles hitting the CCD), and from the
defective pixels (pixels with reduced sensitivity),
with the optimal extraction method of Horne (1986). As the flux
profile, perpendicular to the direction of dispersion, is known, the
Horne's method corrects the pixels which do not match this profile.\\

But in the case of ELODIE, the width of the
orders, perpendicular to the direction of dispersion, is very thin
(3 to 5 pixels). In this case Horne's method misses a significant part
of the cosmics and defective pixels (up to 200 cosmics detectable
for a one hour exposure and $S/N \simeq 50$).
To remove the remaining ``bad'' pixels, we are using three different
and complementary methods on the 1D spectra.
Figure 2 shows an order before and after cosmic rays and defective
pixels removal.\\

\begin{figure}[h]
\resizebox{8cm}{!}{\includegraphics{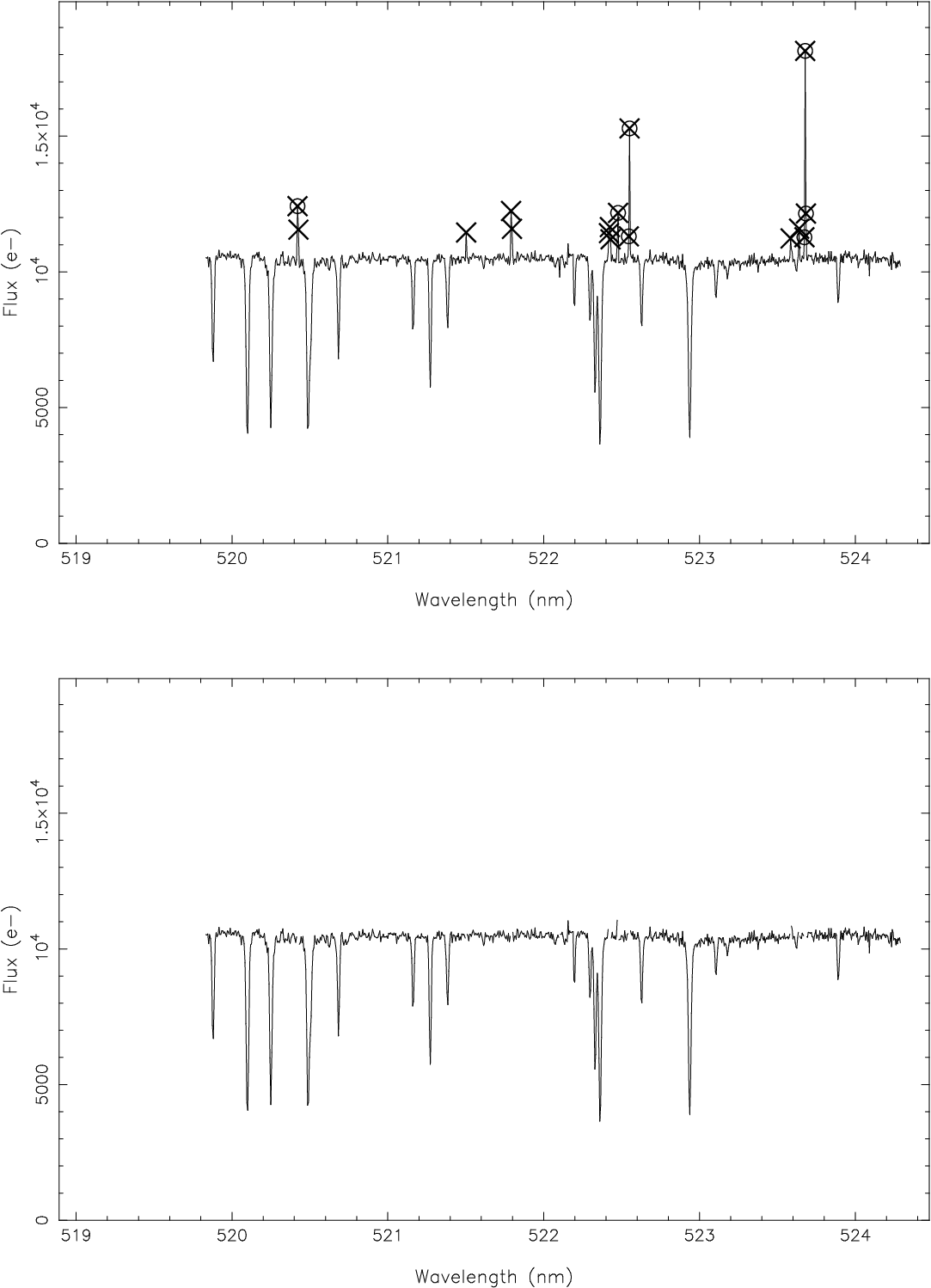}}
\caption{Order 118 of \object{HD 64090} ($T_\mathrm{eff}$ = 5446 K, $\log g$ = 4.45,
[Fe/H] = $-$1.76, S/N = 106)
before and after cosmic rays and defective pixels removal.
Crosses flag cosmics detected by the flux method, circles flag cosmics
detected by the regularity check method.}
\label{}
\end{figure}

\subsubsection{The flux method}
For the stars we are observing ($ 4000 K \le T_\mathrm{eff} \le 6300 K$) and in the 
wavelength interval we are using ($440$ nm to $680$ nm) the lines
are only in absorption. So, every feature that overruns the continuum 
by more than the Poisson photon noise has to be removed. This assumes that the continuum is
well defined.\\

In metal-rich stars the lines are so numerous that the mean and median values of the flux
seriously underestimate the continuum. There are so few true continuum points that they
cannot be successfully fit by a least-squares approach.
Instead, the continuum is estimated by the value of
the $50^{th}$ highest pixel of the order.
A highest pixel was not chosen to avoid the strongest part of the Poisson photon noise.
But in the most metal rich stars, even the value of the
$50^{th}$ pixel slightly underestimates the continuum. Therefore
an index has been defined, which is the number of pixels included between the continuum plus
one sigma  and the continuum minus five sigma,
divided by the total number of pixels of the order.
We have assumed Poisson noise statistics, and taken sigma equal to the
square root of the estimated continuum expressed in number of photo-electrons.
This index is well correlated with the difference
between the true and the estimated continuum, because it gives an
insight of the number and width of the lines which affect the shape
of an order. An order with a few small lines will have a high 
value of its index and a
continuum well estimated, whereas an order with many broad lines will
have a small value of its index, few points belonging to the true 
continuum (often less than $50$) which will be underestimated.
From the estimated continuum
and the index, the limit of the Poisson photon noise is determined
and the pixels with a higher value are removed, giving
them the flag value $-100.0$.\\

This method is especially useful for the cosmic rays that fall
in the continuum (as opposed to those that fall into the lines).

\subsubsection{The regularity check method}
The second method is used to find cosmics, wherever they fall,
and the defective pixels. For every 4 consecutive pixels,
a second order polynomial is fitted on their fluxes, by the least square method . It gives
4 residuals for each pixel.  
The 4 residuals are then quadratically summed to obtain a global
rms deviation for each pixel. If a pixel belongs to the continuum,
it will have a very small rms deviation because of the easiness to fit the 
continuum with a second order polynomial. If the pixel belongs to a
portion of the spectrum containing stellar lines it will have, most of the time,
a larger but still modest rms deviation, because the shape
of the lines is an analytic function, correctly fitted by a second
order polynomial over 4 consecutive points. If a pixel is defective, or
 has been hit by a cosmic, its value is very different from those of its
neighbours. This change is so sharp, that it can't be well fitted
by a second order polynomial, and then the pixel will have a high
residual.\\

Then the mean absolute value of the residuals and their standard deviation $\sigma_r$
over the order are calculated. Two cases are distinguished. 
If the pixel flux is above the estimated
continuum determined by the flux method (Sect. 3.3.1), the pixel is
suspected to have been hit by a cosmic and is removed if its residual 
exceeds the mean by more than $4.6 \sigma_r$.\\
If the pixel flux
is below the continuum, the pixel is a
defective pixel candidate and is removed if its residual 
exceeds the mean by more than $11 \sigma_r$.
This lattest high value of the rejection
coefficient,
comes from the fact, that in a portion containing few narrow lines the rms 
deviation gets close to those of segments containing defective pixels.\\
The elimination procedure is iterated until no more pixels are removed. As in the first method, 
removed pixels are flagged at $-100.0$.  

\subsubsection{The list of defective pixels}
Because of the need to take a high value of the rejection coefficient
in the case of defective pixel candidates,
sometimes some of them are missed.
The third method uses a list of recurrent defective pixels. They were detected
in the stars with broad lines where they are easily distinguished.
This list  is taken as a map of the defective pixels
of the $CCD$. Therefore, all the pixels of this
list are assigned the flag value $-100.0$.\\

This list is unfortunatly somewhat time dependent and is updated at each observing
period. We recall that what is called pixel in the extracted 1D spectrum
is indeed a spectral element involving several true CCD pixels and is dependent on
the position of the spectrum on the CCD.\\

\subsection{Sky subtraction}
The spectrum of the sky is straightened, cleaned of cosmics and defective pixels
and subtracted pixel by pixel from the star spectrum. As the sky subtraction adds noise,
it is suggested to performed it only if the skylight reaches at least 1\% of the starlight.

\subsection{Spectrum convolution}
 The average resolution power of ELODIE (42000) corresponds to a FWHM of 7.14 km.s$^{-1}$
for the line-width. The actual FWHM of spectral lines is the convolution of the intrinsic
line-width by this instrumental line-width. It turns out that the resulting line-width
varies from star to star, but is most of the time between 10 and 13 km.s$^{-1}$
for old F,G, and K stars.  For a young star the FWHM can be much higher.
The resulting resolution
is measured by the FWHM of the cross-correlation function
of the radial velocity determination ($2.35\sigma_{V_r}$, where $\sigma_{V_r}$ is the
standard deviation of the cross-correlation function of the star given by the ELODIE
reduction software).
As the template mask used for the correlation has a large number of lines, the accuracy on the
radial velocity is much smaller than $\sigma_{V_r}$.
For the temperature range of our stars, $\sigma_{V_r}$is usually smaller than $5.5$
km.s$^{-1}$. In this case, to compare spectra at the same
resolution, all the spectra (target and reference stars) were convolved
by a gaussian of standard deviation : $\sqrt{5.5^{2}-\sigma_{Vr}^{2}}$
km.s$^{-1}$.\\ 

Unfortunately for some hot stars ($T_\mathrm{eff} >$ 6100 K),
the $\sigma_{V_r}$ reaches values
higher than $5.5$ km.s$^{-1}$. In this case
TGMET is performed with versions of the
library degraded to lower resolutions.

\begin{figure}[h]
\resizebox{7.2cm}{!}{\includegraphics{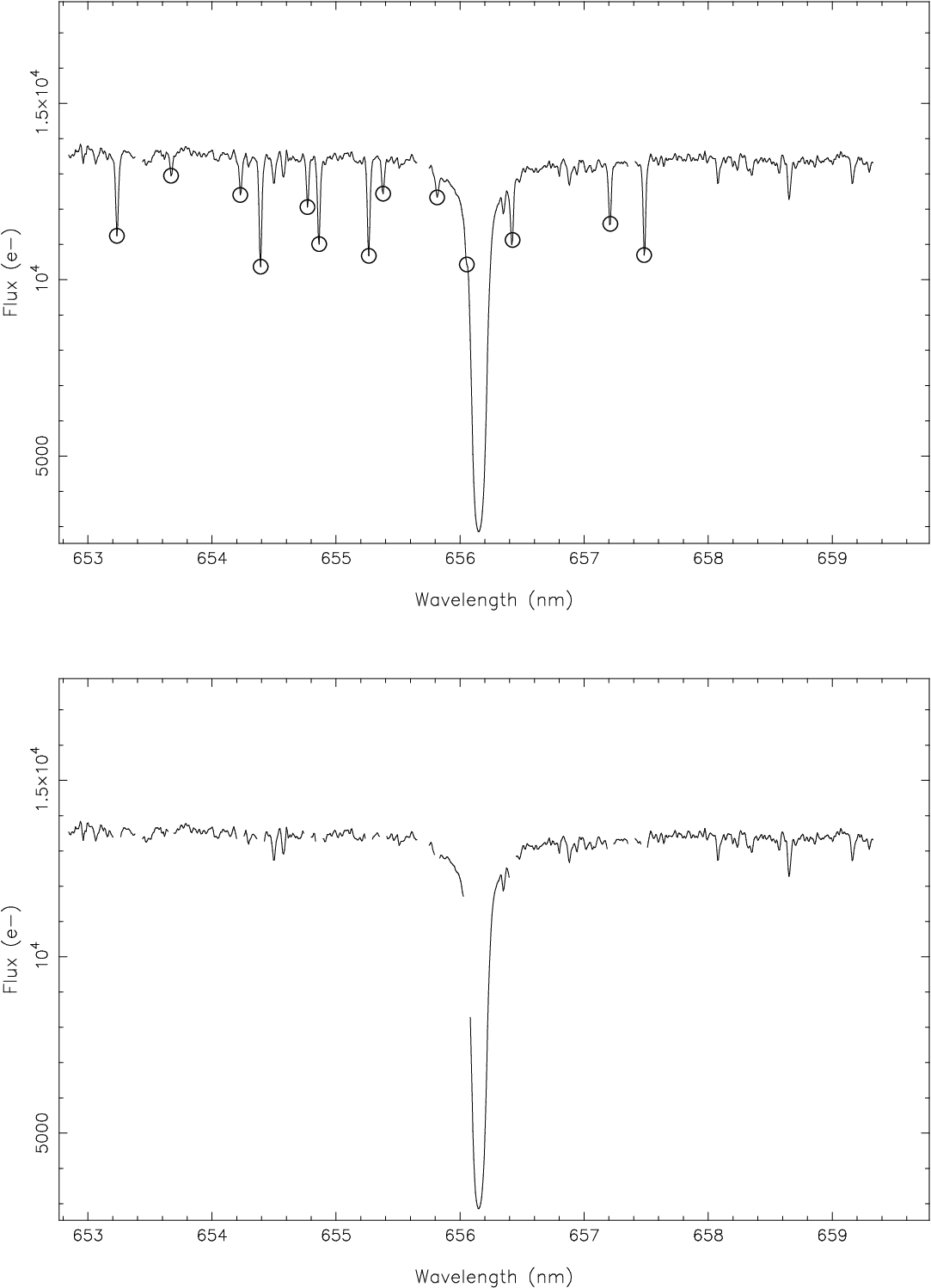}}
\caption{Order $94$ of \object{HD 4306} ($T_\mathrm{eff}$ = 4975 K,
$\log g$ = 2.01, [Fe/H] = $-$2.72,
S/N = 89) before and after atmospheric lines removal. The circles
are showing the wavelengths of the telluric lines as identified in the Rolland's table. 
The broad line at $656.28$ nm is $H_{\alpha}$.}
\label{}
\end{figure}

\subsection{Telluric lines removal}
The stars are observed in different conditions of air mass and humidity
and have different radial velocities with respect to the 
spectrograph. Therefore, positions with respect to the stellar lines
and intensities of telluric lines are changing from one
spectrum to another. They have to be removed to avoid
any earth atmosphere contribution during the comparison between two spectra.\\
 
Telluric lines have been identified using the table of Rowland of
``The solar spectrum'' (Moore et al. 1966). 
Two or three pixels are removed on each side of the 
central pixel of each telluric line, depending on the
equivalent width listed in the Rowland's solar spectrum : smaller or larger than 2 m\AA.
Figure 3 shows the order 94 of \object{HD 4306} before and after telluric lines removal.

\section{ Method of comparison object / library}
The principle of the method is very simple : once the spectrum of an unknown object
is cleaned of instrumental effects, and spurious pixels or lines, it can be adjusted,
order by order, pixel by pixel, 
against several reference spectra by a least square method. The reduced $\chi^2$ of the fit 
is taken as a measure of  resemblance between the target spectrum and
a reference spectrum.
There are two steps for this~:
{\it  wavelength adjustment  and signal amplitude  adjustment.}

\subsection{ Wavelength adjustment}
 The target and a reference star have not the same radial velocity with respect to the
spectrograph, and the first thing to do is to shift the spectrum of the reference star to the 
radial velocity of the object in order to make the absorption lines coincide exactly.
As the difference in radial velocity does not correspond to a whole number of pixels, 
the flux corresponding to a given pixel of the shifted spectrum is not known and 
has to be interpolated. We chose to shift the reference spectrum instead of the contrary 
in order to avoid an interpolation on {\it a too noisy function}. A quadratic Bessel's
interpolation formula was employed.
It makes use of two points before and two points after the considered pixel x~:

$$ f(x)=(1-p)f(x_0) + pf(x_1) + A{p(p-1) \over 4}$$

where:

$$ p = {(x-x_0) \over (x_1-x_0)} $$

$$ x_{-1} < x_0 \leq x < x_1 < x_2$$

and:

$$A = f(x_2) + f(x_{-1}) - f(x_0) - f(x_1) $$
 
This formula is nicely insensitive to noise in the data: if the individual values are affected
by a noise $\epsilon$ the expected noise on the interpolated value is at the most 
$(9/8)\epsilon$.
High order formulae would be amplifying spectrum noise.
 
\subsection{ Mean flux adjustment}

The fluxes of the target and reference stars being usually very different, one has to put 
them on a common scale.  It was decided to move the reference spectrum to the
mean level of the object spectrum. The first idea, based on theoretical considerations, was to
solve this problem by weighted least-squares, finding $u$ such that:

$$  S_k={1 \over n_k-1}\sum_{i\in E_k} \Biggl[{F_{obj}(i,k)-u_k 
F_{ref}(i,k)\over \sigma_{i,k}}\Biggr]^2 $$
 
where: \\
$F_{obj}(i,k)$ is the flux of the target star at pixel i and order k.
$F_{ref}(i,k)$ is the interpolated flux of a reference star interpolated at the same
intrinsic wavelength, as explained in Sect. 4.1.
$E_k$ is the set of pixels of order $k$, for which none of the $F_{obj}(i,k)$ or $F_{ref}(i,k)$ 
are flagged at $-$~100.0 value.
$n_k$ is the number of elements in $E_k$, typically 800 for a single order and
 $\sigma_{i,k}$ is the combined noise affecting the numerator flux difference.\\

Thence $S_k$ is nothing 
else than the reduced $\chi ^2 $ of the fit.
In the
particular case in which a perfect twin of the target star exists 
in the library (no intrinsic
difference, impact of noise only) the expected value of $\chi ^2$ is 1,
with a variance of $2/(n-1)$.
It is of course advantageous to
combine the $\chi^2$ of several orders, because this is equivalent to
increase n, and to decrease the noise $\sqrt{2/(n-1)}$.
For example, with 10 orders, an intrinsic rms difference of only 1.5
per cent can, theoretically, be detected with an observational $S/N$ of 10.  
The implementation needs some 
iterations, because the noise on the numerator depends upon $u$, not known at the beginning.\\

However, after many tests, described in the following section, it was found that good results 
could be obtained with a simpler equivalent technique.
Instead of weighting the terms in the expression of $S_k$ by $({1 \over \sigma_{i,k}})^2$,
we chose to weight them by the response curve
$B_k$ (see Sect. 3.2) which is proportional to ${1 \over \sigma^2}$ 
when the photon noise dominates,
which is the case at $S/N > 10$.
The least square adjustment may then be replaced by finding the value $u_k$ for each order,
which minimizes the expression~:

$$ S^{'}_k= {1 \over n_k-1}\sum_{i \in E_k} (F_{obj}(i,k)-u_k F_{ref}(i,k))^2 B_k(i) $$

giving lower weight to the edges of the order, in the right proportion.   
This leads to a simple formula to determine $u$: \\

$$ u_k = { \sum_{i \in E_k} F_{obj}(i,k) F_{ref}(i,k) B_k(i) 
\over \sum_{i \in E_k} F_{ref}^2(i,k) B_k(i)}  $$
 
Then, the variance is calculated for each order. Figure 4 shows the fit between 
\object{HD 41597} and
\object{HD 40460} (stars with similar parameters)
and Figure 5 shows the fit between \object{HD 41597} and \object{HD 4628}
(stars with different parameters).

\begin{figure}[h]
\resizebox{8cm}{!}{\includegraphics{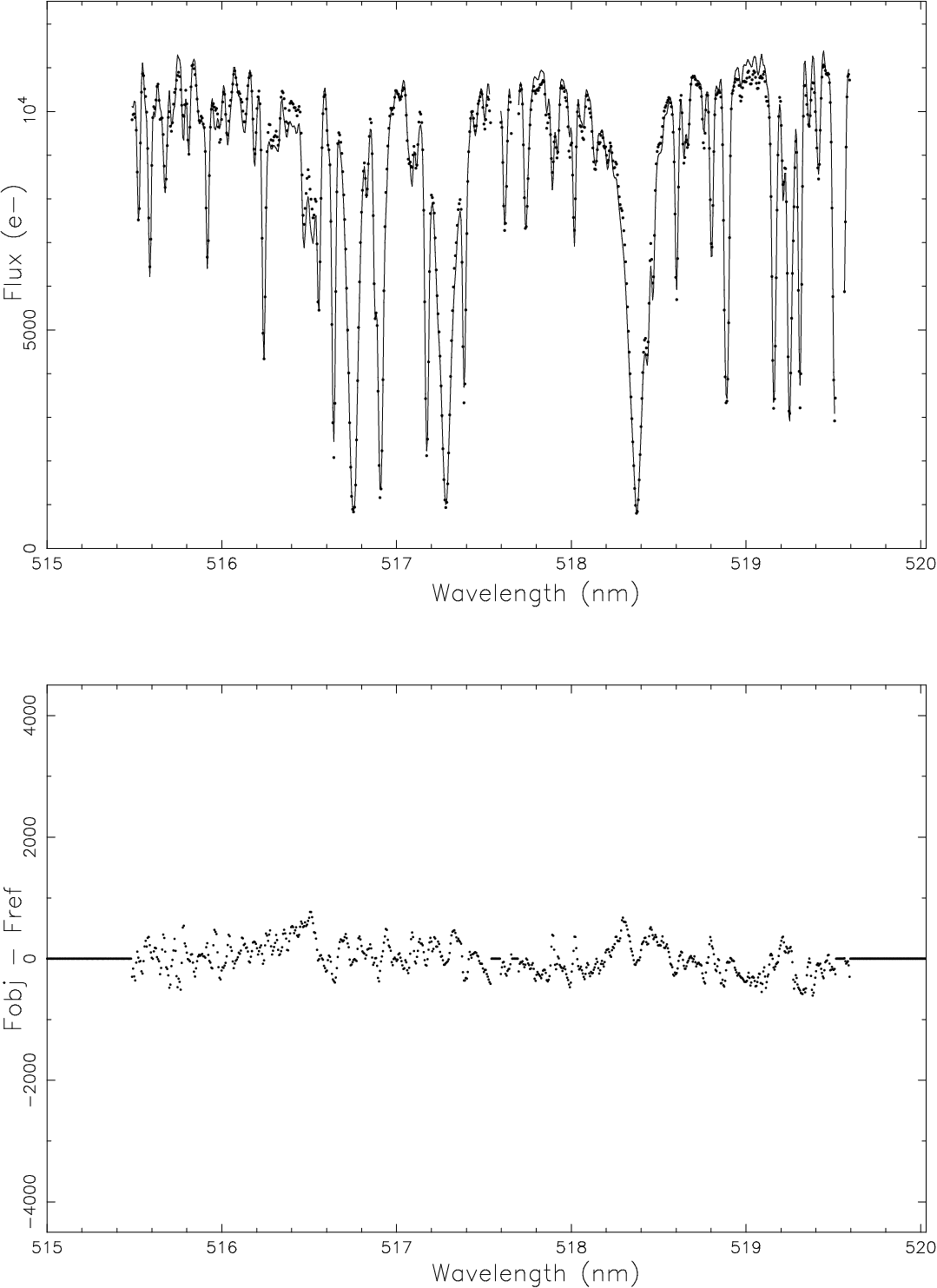}}
\caption{The upper part of the figure shows the fit
between \object{HD 41597}, in points, taken as the target ($T_\mathrm{eff}$ = 4700 K,
$\log g$ = 2.38, [Fe/H] = $-$0.54) and \object{HD 40460} taken as the reference star
($T_\mathrm{eff}$ = 4741 K, $\log g$ = 2.00,
[Fe/H] = $-$0.50). The lower part shows the difference
between their fluxes.}
\label{}
\end{figure}

\begin{figure}[h]
\resizebox{8cm}{!}{\includegraphics{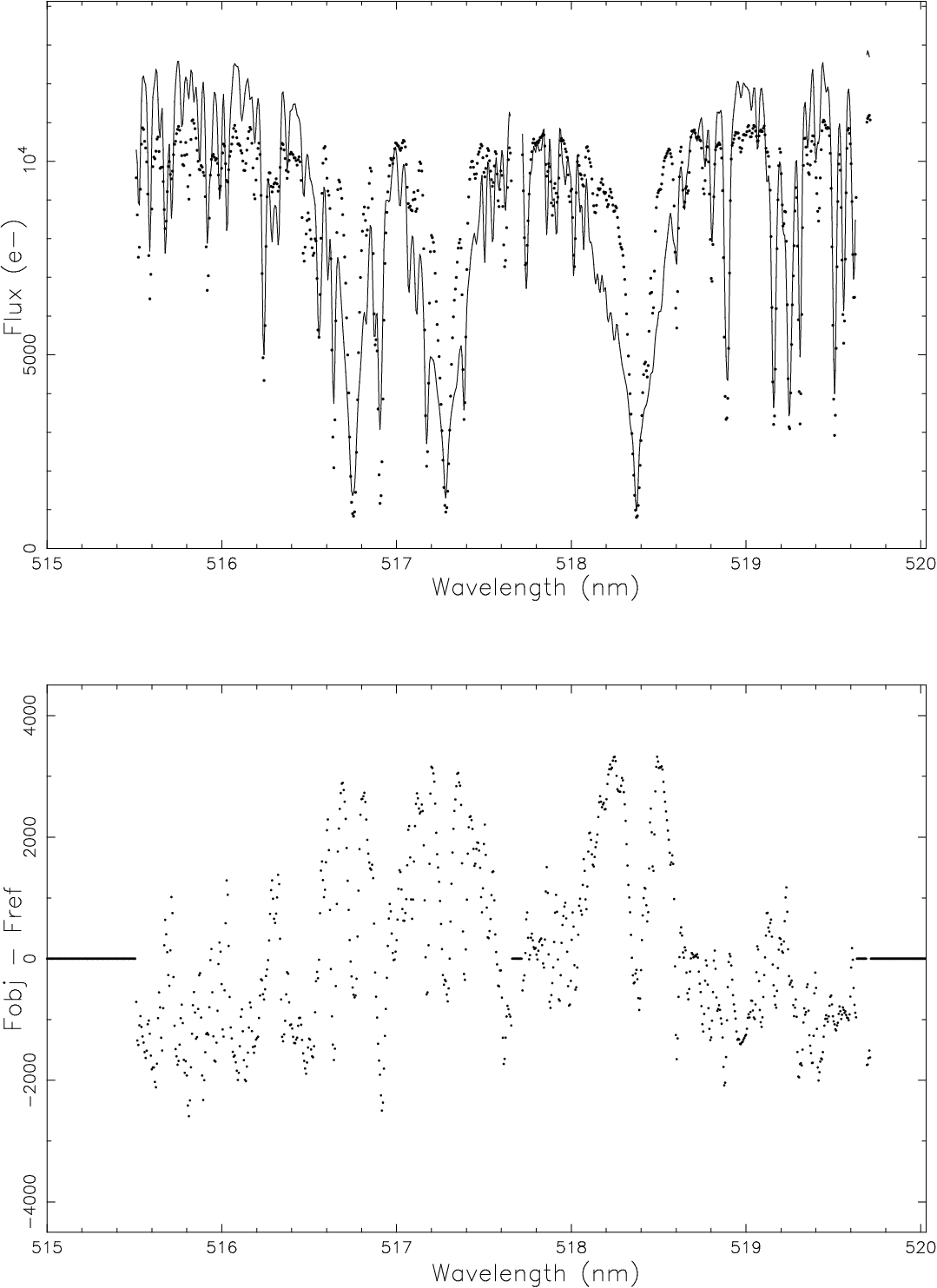}}
\caption{The upper part of the figure shows the fit
between \object{HD 41597}, in points, taken as the target ($T_\mathrm{eff}$ = 4700 K,
$\log g$ = 2.38, [Fe/H] = $-$0.54) and \object{HD 4628} taken as the reference star
($T_\mathrm{eff}$ = 4960 K, $\log g$ = 4.60,
[Fe/H] = $-$0.29). The lower part shows the difference
between their fluxes.}
\label{}
\end{figure}

Now comes the question of properly
combining the information contained in the different orders. This is not a trivial question.
First, there is a large variation of the S/N from the blue to the red orders.
Second the spectral information contained in each order is unequal.\\
 
The first problem is solved by weighting the variance obtained on each order by the the
inverse of its mean flux. The degree of ressemblance between
the target and the reference star is characterized by the expression~:

$$ S_k = {1 \over N-n_b}\sum_{k \in K} {(n_k - 1)S^{'}_k \over <F_k>}$$

with :

$$ N = \sum_{k \in K} n_k$$

where :\\
K is the set of orders kept in the final result(see the next paragraph), 
$n_b$ the number of orders
contained in K and $<F_k>$ the mean flux of the target star for each order.
This expression is in practice equivalent to a reduced $\chi^2$.\\

The second problem is the fact that some orders include strong lines of various
elements, whereas some others are almost free of lines. For example, order 119 including
the Mg I triplet appears to be one of the most significant orders for every star. Orders
94 and 127 containing $H_{\alpha}$ and $H_{\beta}$ are very useful for effective
temperature determinations. Howewer its worth noticing that all orders depend on the
three parameters, which can not be estimated independently. The relative relevance of the
different orders has been studied. In order to select the best orders for each target.
The rms deviation of the fluxes, with respect to the mean flux, caused by
spectral lines plus noise, is computed. The 15 orders for which the ratio of this global
modulation to pure noise modulation is the highest, are kept. 
This set of orders changes from one
target star to another, but the orders 127, 119 and 94 (including respectively $H_{\beta}$,
the Mg I triplet and $H_{\alpha}$) are selected in almost every set.
The number of 15 has been chosen after
a number of ``try and see'' experiments over various S/N.
A larger number of orders introduce noise
instead of information. Figure 6 shows the relative relevance of each order.\\


\begin{figure}[h]
\resizebox{6.1cm}{!}{\includegraphics{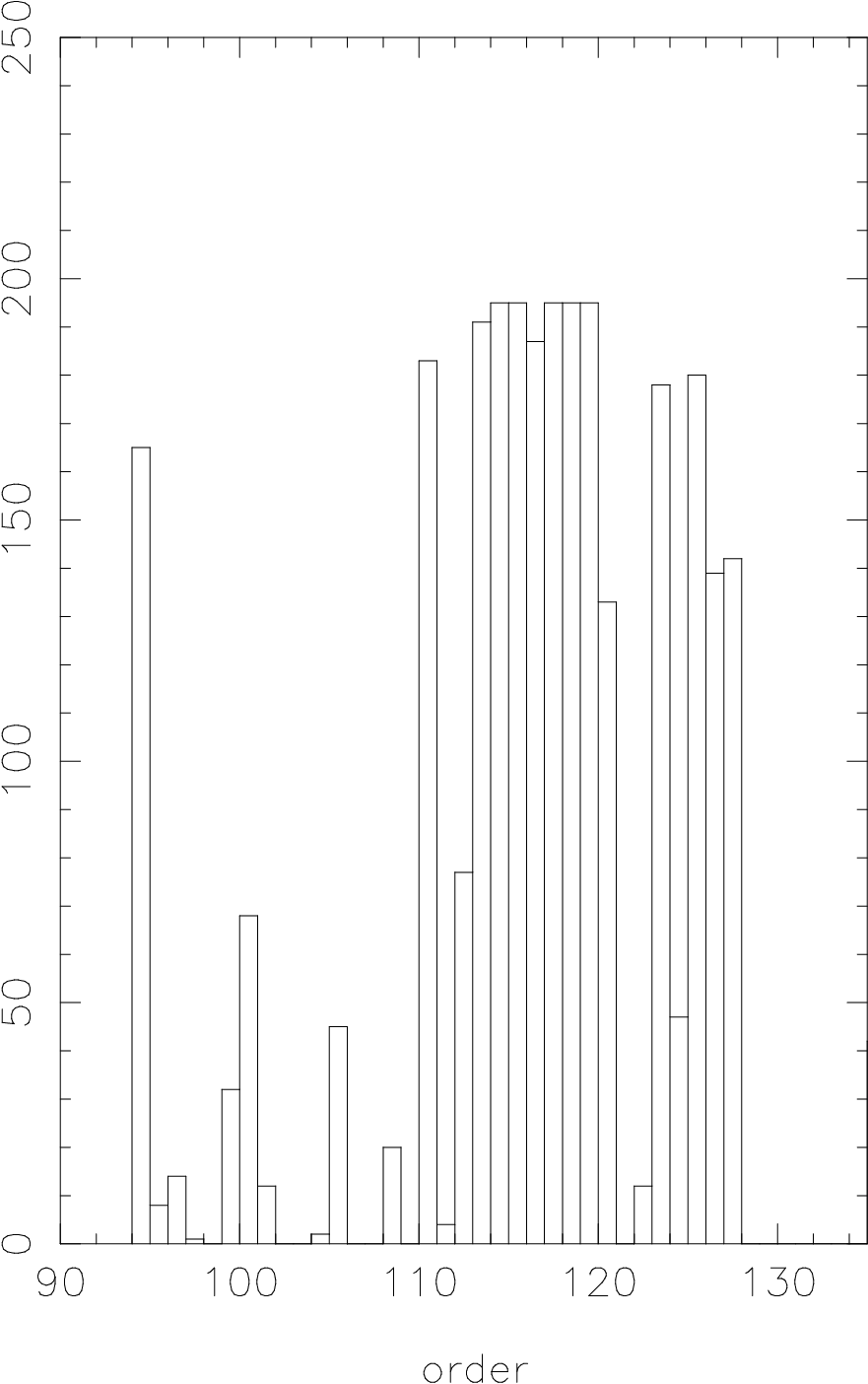}}
\caption{Relative relevance of each order. The test has been performed by
comparing each reference star to the others ($\sim$ 40000 comparisons), by keeping for each comparison 
the 15 orders which give the highest rms and by counting the occurences of each
order in this list.}
\label{}
\end{figure}


For a given target star, the comparison is performed with the 211 reference
stars of the library.

The 10 reference stars showing the lowest reduced $\chi ^2$ are selected.
The solution is given by the weighted mean of the parameters of the reference stars
whose reduced $\chi ^2$ is lower than 1.11 times the lowest one.
The upper limit of $11\%$ has been
empirically defined during the phase of test
(see Sect. 6) and is valuable only for target
stars with $S/N \geq 30$. For smaller S/N the
limit is decreasing (i.e. $1\%$ for S/N = 10).


\begin{figure}[h]
\resizebox{8cm}{!}{\includegraphics{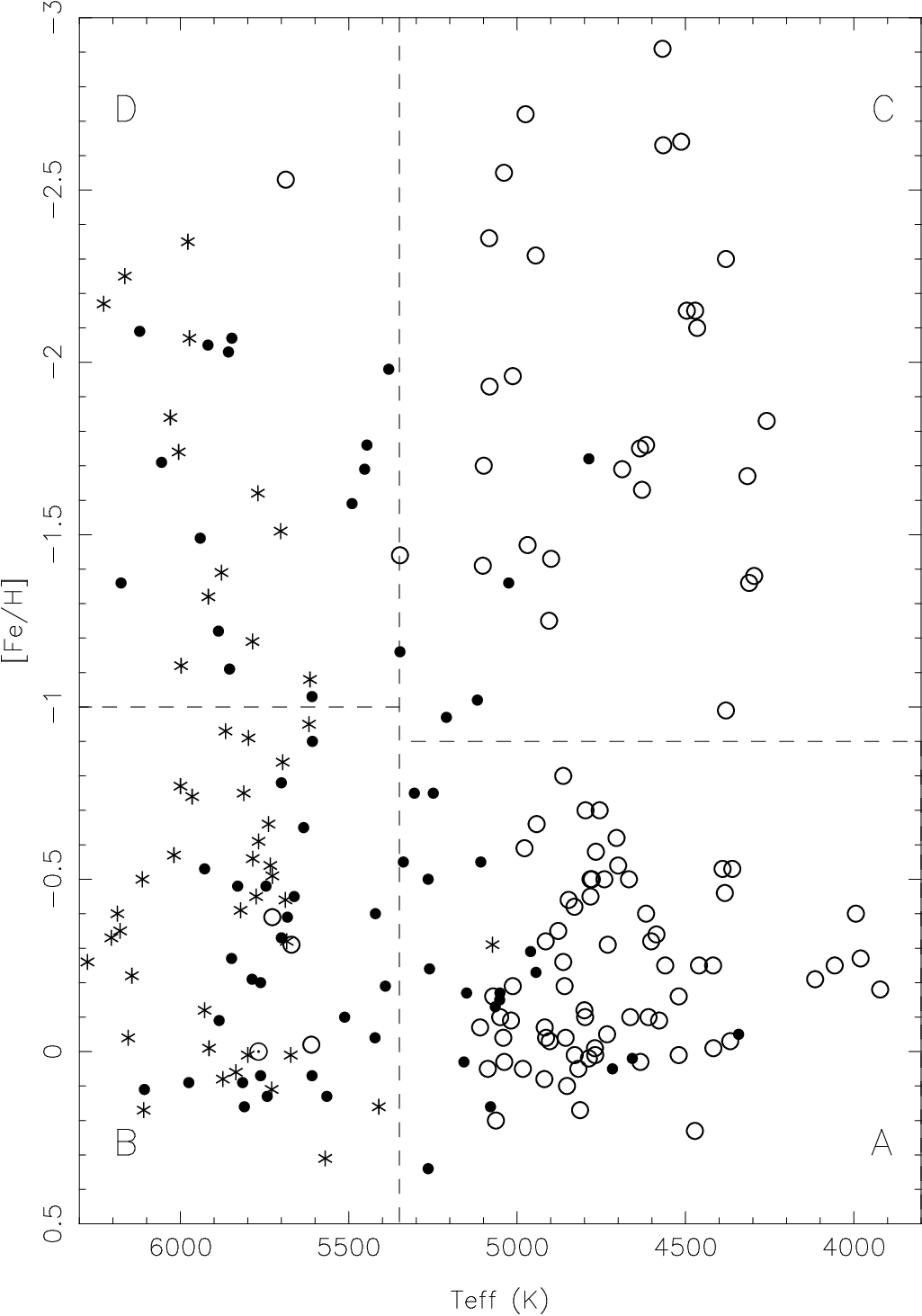}}
\caption{Distribution of the 211 reference dwarfs, sub-giants and giants in the
plane ($T_\mathrm{eff}$, [Fe/H]). Dots represent dwarfs ($\log g \geq 4.2$),
asterisks represent turn-off stars
($4.2 > \log g \geq 3.7$), open circles represent giants ($\log g < 3.7$)
and the Sun is represented by its usual symbol.
The figure is
divided in four zones (from A to D) refered to in Sect. 6.}
\label{} 
\end{figure}

\section{ The library of comparison stars}

The library is made of 211 well-studied stars 
in the temperature range [4000 K; 6300 K]. This temperature interval has been 
chosen because it covers the
full span of stars born from the beginning of the Galaxy, with the elimination of stars
too cool for having a reliable detailed analysis. The metallicity of these
stars is supposed to  
reflect  the metallicity of the interstellar material from which they were formed, tracing
the galactic evolution. In hotter stars, the metallicity might be altered by physical
processes inside the stars, and spectra of cooler stars are complicated and might lead to
erroneous results.\\

The choice of the reference stars and their atmospheric parameters are fully
described in paper II. A large part of this paper is devoted to homogenizing
the parameters because of scattered determinations found in the literature.\\

The success of our method relies on the existence of a
library of high quality, which covers as well as possible the space of the parameters
in the considered temperature interval. Figure 7 shows how dwarfs ($\log g \geq 4.2$), 
turn-off ($4.2 > \log g \geq 3.7$) and
giants stars ($\log g < 3.7$) of the 
library span the plane ($T_\mathrm{eff}$,[Fe/H]). As can be seen, the library
is not homogeneously filled. The holes in the library 
reflect the lack of metal deficient
stars and cool dwarfs, which are more difficult to observe and to analyse.
Of course, in the densest regions of the library the results might be 
statistically better
than in sparse parts.
All the spectra of the library have a signal to noise ratio greater than 35, with a mean
value of 100. We had to keep in the library a few stars with a low S/N to have
enough metal deficient stars, which have  fainter magnitudes.\\

As explained in Sect. 3.5, all the spectra of the library have been convolved to
several resolutions. The lowest one correponds to a FWHM of 13.5 km.s$^{-1}$
and is adequate for most of the F5 to K7 stars. For younger stars with higher rotational velocities
and broadened lines, TGMET uses versions of the library degraded at lower resolutions. 

\section{Testing TGMET }

The first test was to remove one by one each reference star of the library, 
treat it like an anonymous object, then compare the estimated parameters and the
true ones. This test was performed iteratively to adjust the atmospheric parameters of the reference
star as described in paper II. The overall resulting standard deviation of the estimated versus ``true''
parameters for this high S/N sample
is respectively 86 K, 0.28 dex, 0.16 dex for $T_\mathrm{eff}$, $\log g$ and [Fe/H].\\

Several phenomenae influence the results. In metal-rich stars,
the amount of
spectral information is greater than in almost line free stars.
In the same way, stars cooler than
5000 K are losing the Stark profile of the wings of their hydrogen lines.
Therefore the uncertainities
on the determination of the parameters is greater for the cool and
metal-deficient stars than for the
hot and metal-rich stars. To take this effect into account,
the library was divided in four
zones (from A to D, see figure 7) and the mean error was calculated
as well as the dispersion in 
each zone and for each parameter. The results are listed in Table 1.\\

Because our ultimate goal is to use low S/N ($\sim$ 10)
spectra, the spectra
of the library were degraded at S/N = 10 by adding a
random (Poisson) noise.
The mean error and the dispersion in each zone were recalculated. The results are listed in table~2.\\

To be sure that no bias has been introduced with the artificial noise (such as correlation
between pixels), the software was tested with 30 stars of well known parameters
(some of them already belonging to the library), observed willingly at low S/N.
Figure 8 shows the difference between the true and the estimated parameters as a
function of the S/N of the object star.

\begin{table}[h]
\begin{tabular}{| l || l || l | l | l | l |} \hline
Zone                       &  sum &   A  &   B  &   C  &   D  \\ \hline 
number of star             & 211  & 87   & 63   & 33   & 28   \\ \hline \hline
$<\Delta T_\mathrm{eff}>$  & 17   & 11   & 9    & 54   & 11   \\ \hline
$\sigma_{T_\mathrm{eff}}$  & 86   & 74   & 63   & 131  & 87   \\ \hline \hline
$<\Delta \log g>$          & 0.04 & 0.03 & 0.03 & 0.06 & 0.06 \\ \hline
$\sigma_{\log g}$          & 0.28 & 0.25 & 0.20 & 0.43 & 0.28 \\ \hline \hline
$<\Delta$ [Fe/H]$>$        & 0.02 & 0.00 & 0.00 & 0.07 & 0.04 \\ \hline
$\sigma_{[Fe/H]}$          & 0.16 & 0.14 & 0.14 & 0.20 & 0.19 \\ \hline
\end{tabular}
\caption[]{Statistics on the self-concistancy of the library at mean S/N = 100.}
\end{table}

\begin{table}[h]
\begin{tabular}{| l || l || l | l | l | l |} \hline
Zone                       &  sum &   A  &   B  &   C  &   D   \\ \hline 
number of star             & 211  & 87   & 63   & 33   & 28    \\ \hline \hline
$<\Delta T_\mathrm{eff}>$  & 10   & 12   & $-$4   & 50   & $-$9\\ \hline
$\sigma_{T_\mathrm{eff}}$  & 102  & 84   & 78   & 158  & 99    \\ \hline \hline
$<\Delta \log g>$          & 0.03 & 0.00 & 0.04 & 0.08 & 0.05  \\ \hline
$\sigma_{\log g}$          & 0.29 & 0.24 & 0.21 & 0.45 & 0.31  \\ \hline \hline
$<\Delta$ [Fe/H]$>$        & 0.02 & 0.00 & 0.00 & 0.07 & 0.08  \\ \hline
$\sigma_{[Fe/H]}$          & 0.17 & 0.14 & 0.14 & 0.22 & 0.18  \\ \hline
\end{tabular}
\caption[]{Statistics on the self-concistancy of the library at S/N = 10.}
\end{table}


\begin{figure}
\resizebox{8cm}{!}{\includegraphics{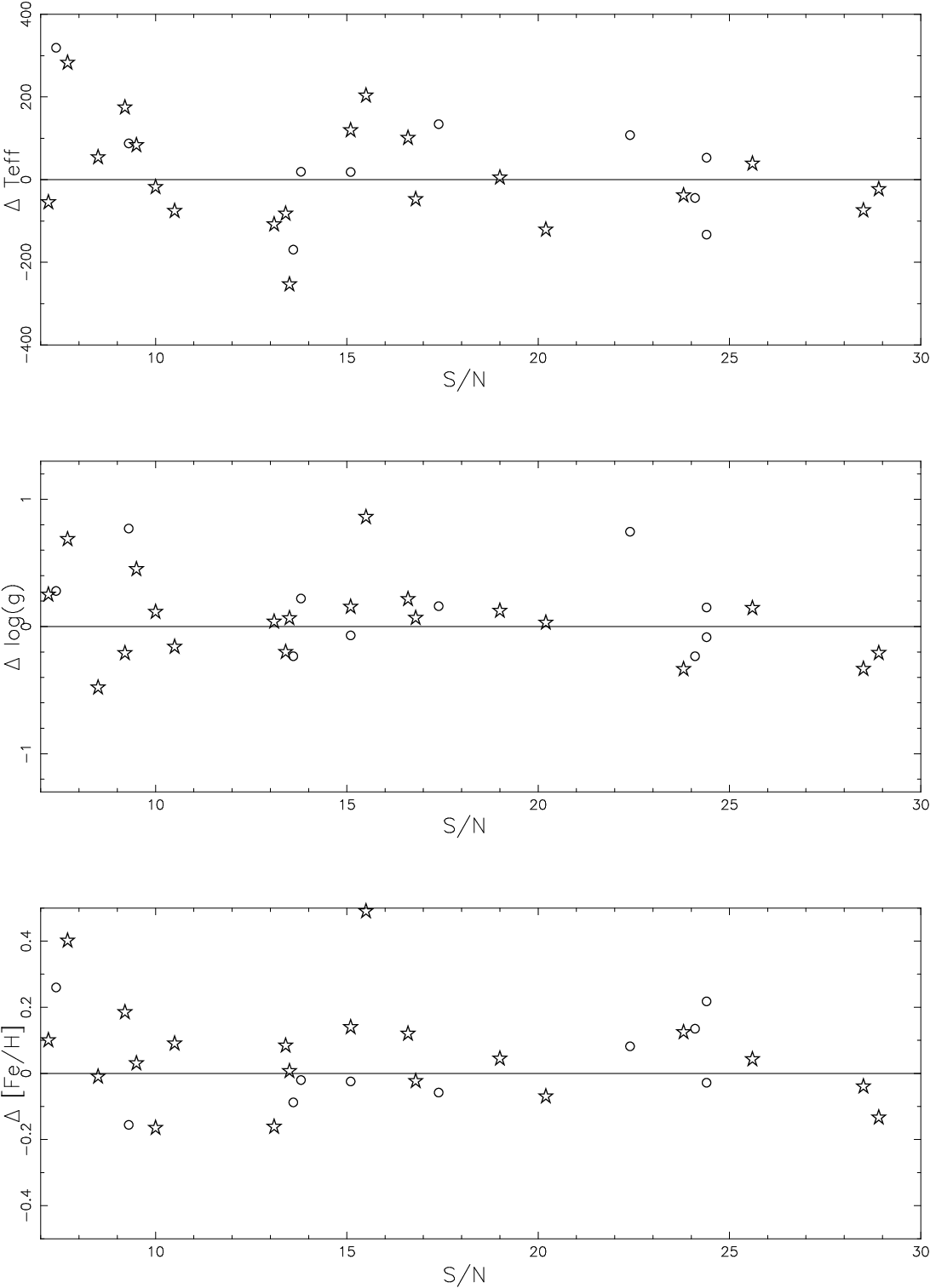}}
\caption{Difference between true and estimated parameters ($T_\mathrm{eff}$ : upper part,
$\log g$ : center part, [Fe/H] : lower part) versus S/N for
30 well-known stars used as test stars. Star symbols flag the stars already
belonging to the library and which have
been reobserved for the purpose of this test at a lower S/N.}
\label{}
\end{figure}

The computing time for TGMET is about 10 min, when run in parallel with the aquisition 
and reduction softwares, on the SUN-sparc 20 station devoted to the observations
with ELODIE.

\section{ Conclusion }

The feasibility of a quick determination, at the telescope, of the fundamental
atmospheric parameters effective temperature, gravity and metallicity of a star
from its spectrum has been demonstrated.
This has been done using spectra supplied by the cross-dispersed echelle spectrograph
ELODIE, mounted at the coud\'e focus of the 1.93~m telecope of the Observatoire
de Haute-Provence. 
These spectra have a spectral resolution of 42 000 and cover a
large spectral range from 380 to 680 nm. We have been 
greatly helped by the already installed ELODIE reduction software, supplying
on-line the radial velocities and the extracted spectra of the stars.\\ 
 
The method consists in comparing the spectrum of the target star with a library
of spectra taken with the same spectrograph, at a typical S/N of 100. Such a 
library has
been assembled , and presently contains the spectra of 211 stars, selected for
covering the parameter values encountered in stars belonging to the halo, the
thick disk
and the old thin disk of the Galaxy. Namely the parameter intervals are
[4000K, 6300K],
[0.6, 4.7] and [$-$2.9., 0.35] respectively in $T_\mathrm{eff}$, $\log g$
and [Fe/H].\\
 
A great amount of care has been put into the elimination of all features not
intrinsic to
the stars before attempting any comparison between their spectra. In particular
we have removed
the modulation of the spectra by the blaze of the orders (which still exists even 
if all spectra are 
taken with the same spectrograph because of the radial velocity shifts), the
telluric lines
and the hits by cosmic rays.  It was also necessary to bring all spectra
to a common resolution
to avoid differences due to stellar rotation or macroturbulence or to a slight 
difference 
in focus during different observing runs.
The degree of resemblance between the spectra of the target and those of the
library was 
quantified by a maximum likelihood approach, involving a weighted least-square 
fit. The parameters 
of the target star are determined by taking a weighted mean of the the parameters
of the most resembling reference stars.\\

The accuracy of the method, with the existing library, has been tested. On the 
average, the accuracy 
turns out to be 86 K in $T_\mathrm{eff}$, 0.28 dex in $\log g $, and
0.16 dex in [Fe/H] for
a target spectrum with S/N = 100. This accuracy critically depends upon the 
number of spectra in the
library and the quality of the parameters collected for each reference object
in the literature.
A second paper ( Soubiran et al. 1998) deals with this point in more detail,
probing the consistency
of the library by internal comparison of all the spectra of the library.\\
 
Our fundamental stellar parameter extraction software (named TGMET for
Temperature, Gravity, METallicity) is available to all ELODIE users at Observatoire
de Haute-Provence. It is online, so that these parameters may be derived during
observing runs.
Its user-guide is
also available on the WEB at http://www.obs-hp.fr.\\
 
A similar software is planned for the new spectrograph FEROS at La Silla (ESO).\\
 
Future improvements are planned and already under way. The most obvious
ones are to fill the sparsely
populated zones of the library, and in redetermining the fundamental parameters
of the stars
which have been spotted as doubtful in the consistency check. Another one is 
to extend the library of reference stars up to F0, at the request of other observers.
In particular the asteroseismologists working on the
preselection of target stars for the spatial experiment COROT with ELODIE
need atmospheric parameters for stars from A to G type.
We plan also to improve
the algorithm determining the parameters, by taking more rigorously into
account the sensitivity
of each pixel of the spectrum to a variation of the parameters. This will
be particularily useful
for very metal-poor stars, for which only very few  places in the spectrum
contain useful information.
An exciting extension to our software would be to derive abundance ratios
in addition to metallicity. For example we could try to determine the
alpha-elements to iron ratio, which has been shown to present a scatter larger
than believed before (King 1997, Carney et al. 1997, Nissen \& Schuster 1997).
This ratio is particularly important, as the Mg I green triplet and the molecular
band MgH are dominant gravity features. Other abundance ratios of interest as
[Ba/Fe], [Sr/Fe] or [Na/Fe] are attractive targets.

\begin{acknowledgements}
We are very grateful to Didier Queloz for providing his valuable software for
ELODIE's spectra reduction
and for fruitful discutions. We want to thank Alain Vin for his constant help during the
observational runs. We thank the astronomers who let us use their spectra
to fill the library. Many thanks to M.-N. Perrin, C. Sneden and P.E. Nissen for very
useful comments and suggestions for improving the manuscript.
\end{acknowledgements}

\end{document}